\documentstyle[prl,aps]{revtex}
\begin{document}
\draft
\twocolumn[\hsize\textwidth\columnwidth\hsize\csname
@twocolumnfalse\endcsname

\title{Anomalous spin and charge dynamics of the t-J model at
low doping}

\author{R. Eder, Y. Ohta, and S. Maekawa}

\address{Department of Applied Physics, Nagoya University,
Nagoya 464-01, Japan}

\date{\today}
\maketitle

\begin{abstract}
We present an exact diagonalization study of the dynamical spin and
density correlation functions in small clusters of $2D$ $t$$-$$J$ model
for hole dopings $\leq$$25$\%. Both correlation functions show
a remarkably regular, but very different scaling with both
hole density $\rho_h$ and parameters $t$ and $J$:
the density correlation function is most consistent with that of
condensed Bosons in a band of width $\sim$$t$, the spin correlation
function with that of fermions in a band of width $\sim$$J$.
We show that the familiar spin bag scenario explains these results
in a natural way.
\end{abstract}


\vskip2pc]
\narrowtext

The identification of a simple
`effective theory' for the doped Mott-Hubbard insulator which is
capable of resolving the numerous anomalies of cuprate
superconductors remains an intriguing problem.
The simplest model which may be expected to contain the key features
of these systems is the
$t$$-$$J$ model:
\[
  H =
 -t \sum_{< i,j >, \sigma}
( \hat{c}_{i, \sigma}^\dagger \hat{c}_{j, \sigma}  +  H.c. )
 + J \sum_{< i,j >}[\;\bbox{S}_i \cdot
 \bbox{S}_j
 - \frac{n_i n_j}{4}\;].
\]
The $\bbox{S}_i$ are the electronic spin operators,
$\hat{c}^\dagger_{i,\sigma} =c^\dagger_{i,\sigma}(1-n_{i,-\sigma})$
and the sum over $<i,j>$ stands for a summation
over all pairs of nearest neighbors. \\
For hole densities $\rho_h$$>$$0.3$ the ground state of the
model seems to represent a fairly conventional Fermi liquid
with a particle-hole like spin and density excitation
spectrum\cite{Tohyama,lowdope}.
In this work we show that the situation is
drastically different for $\rho_h$$\leq$$0.25$:
here the density excitations roughly resemble
those of condensed Bosons with characteristic energy $t$,
the spin excitations still are consistent
with Fermions, the characteristic energy, however, now is $J$.
We show that these results are naturally explained
within the familiar spin bag\cite{Schriefferetal}
or string\cite{Shr1,Trugman,Maekawa,DagottoSchrieffer} scenario.\\
For the standard $16$ and $18$-site clusters we
used the Lanczos algorithm to compute the dynamical
spin (SCF) and density (DCF) correlation functions:
\[
C_\alpha(\bbox{q},\omega)
= \frac{1}{\pi} \Im \langle \Psi_0 | O_\alpha^\dagger(\bbox{q})
\frac{1}{ \omega - ( H- E_0 ) - i\epsilon} O_\alpha(\bbox{q})
 | \Psi_0\rangle.
\]
Here $|\Psi_0\rangle$ ($E_0$) denotes the ground state
wave function (ground state energy),
for the operator $O_\alpha(\bbox{q})$
we choose the Fourier transform of either
density operator $(n_{i,\uparrow}$$+$$n_{i,\downarrow})$
($\alpha$$=$$d$) or spin operator
$(n_{i,\uparrow}$$-$$n_{i,\downarrow})$ ($\alpha$$=$$s$).
Fig.\ref{fig1} shows the DCF {\em divided by the number of holes},
$n_h$, for $n_h$$=$$1,2,3,4$. Except for the $\omega\rightarrow 0$
parts at $(\pi/2,0)$ and $(\pi/3,\pi/3)$ this is a universal
function. Deviations are strongest for
a single hole, but even there the characteristic shape
is already present.
Fig.\ref{fig1} thus first of all demonstrates a high degree of
continuity over the entire range of dopings considered;
quite obviously the essential physics is already realized
for a single hole in an antiferromagnet.
Next, the scaling of the spectra with hole concentration
over their entire width is clearly inconsistent
with particle-hole excitations in a Fermion system: at least
the high-energy parts, where transitions from
deep below to high above the Fermi energy $E_F$ would
contribute, should be unaffected by a change of particle
density. Instead, the DCF could be modelled roughly by
Bosons of density $\rho_h$ which are
condensed into the lowest state of the noninteracting
band with the free electron dispersion $\epsilon_{\bbox{q}}$;
such a system would have a DCF of the form
$C_d(\bbox{q},\omega)$$=$$\rho_h \cdot
\delta( \omega - \epsilon_{\bbox{q}})$.
That the characteristic energy of the DCF indeed is $t$ is seen
in Fig. \ref{fig2}, which compares the DCF for different
$t/J$ and demonstrates that the
positions of the dominant `peaks' remain unaffected by a change
of $J$. Fig. \ref{fig2}
 also shows that the weight of the $\omega\rightarrow 0$ part
decreases with $t/J$, whereas the weight of the remainder
other parts increases; together with the deviations in the
scaling behaviour with $n_h$ this suggests a two-component
interpretation of the DCF.\\
We proceed to the SCF, shown in Fig. \ref{fig3} for
$n_h$$=$$2,3,4$.
The spectra for $3$ holes thereby are averaged
over the $4$ degenerate ground states (the $3$ hole ground states
have momentum $(2\pi/3,0)$ in the $18$ and $(\pi/2,\pi/2)$
in the $16$-site clusters). With the exception of
$(\pi,\pi)$, the spectra are fairly independent of doping,
changes occur mostly at the low-energy end of some
of the spectra. Such a dependence on particle number
is reminiscent of
the particle-hole excitations in a Fermion
system: the spectral weight of a given transition
is expressed in terms of the Fermionic occupation numbers
$n_{\bbox{k}}$ as $n_{\bbox{k}+\bbox{q}} (1-n_{\bbox{k}})$
and is affected only if either initial or final state
cross the chemical potential. Here it should be noted that
due to the coarse $\bbox{k}$-mesh of the $2D$ clusters
(as well as the averaging process for $3$ holes)
one can not expect a well defined $\bbox{k}_F$
which scales continuously with particle density.
Instead a countinuous increase of the hole occupation numbers
$n_{\bbox{k}}$ at the `Fermi momenta'
and hence a continuous increase/decrease of peak intensities
will be realized. As for the energy scale
Fig. \ref{fig4} shows the SCF for various
$t/J$. When frequencies are measured in units of $J$, the
peak positions obviously remain largely unaffected by a
change of $t$, so that the characteristic energy
of the SCF is $J$.\\
Summarizing the numerical results one may say that
both SCF and DCF scale with $n_h$ and $t/J$
in remarkably regular but completely
different ways. Let us stress for clarity
that the different doping dependence of the
{\em integrated} weight with nonvanishing momentum
transfer is trivial. By the $f$-sum rule the
integrated spectral weight of both correlation functions
equals the number of electrons, $n_e$.
For the DCF this value is almost exhausted
by the $\bbox{q}$$=$$0$ spectrum,
which contributes $n_e (n_e/N)$ (with $N$ the number of sites).
For the SCF this contribution either vanishes or is a tiny
$1/N$, so that
the integrated spectral weight for finite $\bbox{q}$
is $\sim$$n_e$ for the SCF but $\sim$$n_h$
for the DCF. This argument, however does not
explain e.g. the scaling with $n_h$ of the entire DCF
spectra. This is not only inconsistent
with the particle hole excitations in a Fermi liquid
but also with the Luttinger liquid
realized in the $1D$ model\cite{Tohyama,BaresBlatter}. The low
doping phase in $2D$
thus obviously exhibits qualitatively new physics.\\
To gain additional insight, Fig. \ref{fig5}
compares the SCF for $t/J$$=$$2.5$ and $t$$=$$0$,
i.e.,  mobile and static holes. For static holes there is a
band of diffuse excitations, which roughly follow
the characteristic spin wave dispersion; their diffuse nature is
obviously due to the scattering from the holes, which in this case
merely act as impurities.
For mobile holes spin-wave excitations
near $(\pi,\pi)$ seem to persist with reduced spectral weight
and a wider gap at $(\pi,\pi)$; the qualitatively new feature are
low energy peaks throughout the Brillouin zone which may
naturally be associated with particle-hole
transitions in the coherent band for mobile holes.
This suggests a two-component interpretation for the SCF, namely
low-energy particle-hole excitations plus spin-wave like
excitations near $(\pi,\pi)$.
This is also supported by the opposite
$t/J$-dependence of these two components (Fig. \ref{fig4}):
whereas the weight of the `spin wave' parts decreases with
increasing $t/J$ (consistent with the more efficient
degradation of antiferromagnetic correlations
by more mobile holes)
that of the particle-hole-excitations increases or remains unchanged.\\
Further information is obtained by studying
the impact of  $O_s$ and $O_d$ on
the electronic momentum distribution (EMD)
$n(\bbox{k})=\langle \hat{c}_{\bbox{k},\sigma}^\dagger
\hat{c}_{\bbox{k},\sigma}\rangle$. To that end
we compute first the ground state EMD,
 $n_{0}(\bbox{k})$, and second the EMD
$n_{\bbox{q}}(\bbox{k})$ for the state
$(1/n)O_\alpha(\bbox{q})
|GS\rangle$ (where $n$ is chosen to normalize the state to $1$).
Table \ref{tab1} shows the difference
$\Delta_{\bbox{q}}(\bbox{k})$$=$$n_{\bbox{q}}(\bbox{k})
$$-$$n_{0}(\bbox{k})$
for $\bbox{q}=(\pi,\pi)$, as well as
$n_{shift}(\bbox{q})
= \sum_{\bbox{k}} |\Delta_{\bbox{q}} (\bbox{k})|$. The
latter quantity may be interpreted as the number of
electrons shifted in $\bbox{k}$-space, its value for free
particles is $1$.
Table \ref{tab1} reveals a clear difference between
density and spin operator: whereas
$O_d$ induces a substantial shift of electrons in
$\bbox{k}$ space and always has $n_{shift}$ close to its
free-particle value of $1$, $O_s$ affects $n(\bbox{k})$
to a much lesser
degree. Via the kinetic energy sum rule
\begin{equation}
E_{kin} = 2 \sum_{\bbox{k}} \epsilon_{\bbox{k}}
n(\bbox{k}),
\label{kin}
\end{equation}
(with $\epsilon_{\bbox{k}}=-2t (\cos(k_x) +\cos(k_y))$),
this result on one hand immediately explains the
different energy
scales of SCF and DCF: $O_d$ substantially changes
$E_{kin}$$\sim$$t$, $O_s$ leaves it essentially unchanged.
On the other hand, this only complicates
the puzzle about the doping dependence of the DCF:
quite obviously
the electrons react to the density operator as if
they were free particles, hence one naturally would expect
a Fermion-like particle-hole spectrum for the DCF,
in contrast to the numerical result.\\
We now want to show that all results obtained
so far can be resolved in a simple and natural way
if one adopts the familiar string or spin bag picture.
Thereby we consider holes moving in (and hence coupled to)
a `background' of
antiferromagnetically correlated
spins. Our key assumption is that the spin background
carries excitations
which are independent of the hole system, so that their
momentum is
not `visible' in the EMD (compare Table \ref{tab1})
and hence does not increase the kinetic energy.
Natural candiates are the remnants of the short-wavelength
spin waves, as would be suggested by Fig.\ref{fig5}.
We next assume that the relevant hole states are described
by operators of the type

\begin{eqnarray}
\tilde{c}_{\nu,\bbox{k},\uparrow}
&=&\alpha_\nu(\bbox{k}) \hat{c}_{\bbox{k},\uparrow}
+ \sum_{\bbox{k}'} \sum_{\sigma,\sigma'} \beta_\nu^{\sigma,\sigma'}
(\bbox{k},\bbox{k}')
\hat{c}_{\bbox{k}',\sigma} S^{\sigma'}_{\bbox{k}-\bbox{k}'}
\nonumber \\
&+& \sum_{\bbox{k}',\bbox{q}} \sum_{\sigma,\sigma',\sigma''}\gamma_\nu^
{\sigma,\sigma',\sigma''}(\bbox{k},\bbox{k}',\bbox{q})
\hat{c}_{\bbox{k}',\sigma} S^{\sigma'}_{\bbox{q}}
S^{\sigma''}_{\bbox{k}-\bbox{k}'-\bbox{q}}
\nonumber \\
&+& \dots,
\label{bag}
\end{eqnarray}

where $S^{\sigma}_{\bbox{q}}$ denotes the electronic spin operator.
Equation (\ref{bag}) describes a hole which has transferred a
part of its
momentum to a variable number of spin excitations;
it has many degrees of freedom
(as reflected by the many parameters $\alpha$, $\beta$, $\gamma$
$\dots$) so that there will be a large number of bands, which
presumably form the extended incoherent continua present in the
single particle spectral function\cite{Maekawa,Szczepanski}
The bands are labelled by the
index $\nu$, $\nu$$=$$0$ denotes the
`quasiparticle band'\cite{Szczepanski} split off from
the bottom of the continuum.
For this quasiparticle band the validity of the spin bag
description has been verified previously by explicit
numerical check\cite{DagottoSchrieffer,spinbags}.
The driving force behind the momentum transfer to
spin excitations
is gain in kinetic energy: scattering a hole e.g. from
$\bbox{k}$$=$$(\pi/2,\pi/2)$ to
$\bbox{k}'$$=$$(\pi,\pi)$ and transferring the excess
momentum to a spin excitation reduces the kinetic
energy $\sim$$t$, but requires only exchange energy
$\sim$$J$$<$$t$. We can conclude that
in the low lying spin bag states
the bare hole predominantly occupies momenta near $(\pi,\pi)$:
via (\ref{kin}) the depletion of $n(\bbox{k})$
on these momenta lowers the kinetic energy most efficiently.
This is corroborated by
the numerical result\cite{Ding} that
the addition of e.g. a single hole with total momentum
$(\pi/2,\pi/2)$ reduces $n(\bbox{k})$ strongest
near $(\pi,\pi)$.
A rough estimate for the degree of
admixture of spin excitations is the quasiparticle
weight $Z$: the simplest estimate would be
$Z$$=$$|\alpha_0(\bbox{k_F})|^2$;
$Z$$\simeq$$0.3$\cite{Dagottoetal} thus suggest a rather
strong admixture of spin fluctuations.\\
We now assume that in the $n_h$ hole-ground state the holes
occupy the $n_h$ lowest states of the $\nu$$=$$0$ band, i.e.
rigid-band filling of the quasiparticle band; this is
consistent with numerical results\cite{rigid}.
Since the dispersion of the $\nu$$=$$0$
band strictly scales
with $J$\cite{Szczepanski}, we may conclude from (\ref{kin})
that the distribution
of the bare holes in $\bbox{k}$ space is essentially the same
for all $\nu$$=$$0$ states. In other words, the coefficients
$\beta$, $\gamma$ in (\ref{bag}) depend strongly on the
`bare hole' momentum $\bbox{k}'$ but only weakly on the
total momentum $\bbox{k}$. We can conclude that when
$n_h$ holes are filled into the $\nu$$=$$0$ band,
the hole occupation of
e.g. $(\pi,\pi)$ should be $n_h$ times that for a single hole;
that this indeed is the case is shown in Table \ref{tab2} which gives
the hole occupation, $\bar{n}(\bbox{k})$$=$$ (1/2)\cdot
\sum_\sigma\langle c_{\bbox{k},\sigma}
c_{\bbox{k},\sigma}^\dagger \rangle$,
for momenta near $(\pi,\pi)$ in the ground state with different
$n_h$.\\
With this picture of the ground state in mind,
we can distinguish two types of excitations:
there can be `particle-hole excitations' within the $\nu=0$
band or an
excitation of the `internal degrees of freedom' of a spin bag,
where the final state has $\nu'$$\neq$$0$.
An important point is that $O_s$ and $O_d$
transfer momentum to a spin bag in a very different way:
$O_s$ changes the momentum of the bag by adding
(or removing) a spin excitation, whereas $O_d$
can transfer its momentum only to the bare hole itself. $O_s$
consequently does not change the distribution of holes
in $\bbox{k}$-space appreciably
whereas $O_d$ necessarily must do so (see Table \ref{tab1}).
Hence, $O_s$ is essentially limited to transitions within
the $\nu$$=$$0$ band, so that its spectrum
is Fermionic and scales with the quasiparticle bandwidth
$J$. On the other hand, $O_d$ induces transitions
from the $\nu$$=$$0$
band to `bands' in the continuum, with an excitation energy
$\sim$$t$ which usually far exceeds the bandwidth $\sim$$J$.
Then, particle-hole transitions of Fermions
between bands of width $W$, which are separated in energy by
$E$$\gg$$W$, will give the contribution
$C(\omega)\simeq \rho_f \delta( \omega - E)$,
to the correlation function,
where $\rho_f$ is the density of Fermions in the lower
(partially occupied) band.
This immediately explains the {\em apparently} Bosonic doping
dependence
of the DCF, which thus originates from the
possibility to excite high-energy degrees of freedom
of the structured spin bag quasiparticles.
Only for small momentum transfer the increase in
kinetic energy is small, so that $O_d$ also can
generate particle-hole transitons in the $\nu$$=$$0$ band,
where restrictions due to the Pauli principle apply, hence
the deviations from the scaling of the DCF
with $n_h$ in the low energy region.
The spin bag scenario thus provides a natural
explanation for the unusual scaling behaviour of the
different correlation functions; it is supported by
detailed consistency with a substantial body of
numerical evidence. For completeness we note that
slave-boson mean-field theories\cite{sbmf}
describe the ground state of the $t$$-$$J$ model as
a product of condensed Bosons in a band of width $\sim$$8t$
and Fermions in a band of width $\sim$$4J$. With the additional
assumption that the density operator acts only on the
bosons, the spin operator only on the Fermions, this ground
state clearly would have an excitation spectrum which is consistent
with the numerical results; the justification of this assumption
as well as the possible agreement with details of the
cluster results remains to be clarified.\\
In summary, we have studied the dynamical spin and
density correlation function for the $2D$ $t$$-$$J$ model
near half-filing.
Whereas these correlation functions should be closely
related in a Fermi liquid,
we found them to differ substantially for this strong correlation
model:
the density correlation function has a Boson-like dependence on the
hole density and the hopping integral $t$ as its characteristic
energy scale, the doping dependence of
the spin correlation function is consistent with
Fermions and it has the exchange constant $J$ as energy scale.
While the remarkably systematic scaling of the correlation functions
suggests the existence of a simple `effective theory' for the
excitation spectrum, the familiar particle-hole picture thus
is clearly insufficient.
The familiar spin bag picture then provides a promising
framework for such an effective theory:
the strong dressing of the hole
with spin fluctuations  and the resulting complex
internal structure of the spin bag-type quasiparticles
lead to a qualitatively new type of
excitations, namely the excitation of internal degrees of
freedom of the quasiparticles.
Spin and density operator differ markedly in their
ability to excite the various degrees of freedom of the
spin bag liquid, hence their very different spectra.
While details need to be worked out, it seems obvious that
the existence of such new types of excitations, as well as
the apparently very different response of the spin bag
liquid to `spin-like' and `charge-like' perturbations
may lead to experimentally observable
anomalies e.g. in high-temperature superconductors.\\
Financial support of R. E. by the Japan Society for the Promotion
of Science is most gratefully acknowledged.
\figure{DCF divided by $n_h$
for various $n_h$ ($t/J=2.5$, Lorentzian broadening
$\epsilon$$=$$0.2t$).
\label{fig1}}
\figure{DCF for $2$ holes in
$16$-sites for different $t/J$ ($\epsilon$$=$$0.2t$).
\label{fig2}}
\figure{SCF for different $n_h$; the $(\pi,\pi)$ spectra are
multiplied by $0.2$ ($t/J=2.5$, $\epsilon$$=$$0.5J$).
\label{fig3}}
\figure{SCF for $2$ holes in $16$ sites for different $t/J$; the
$(\pi,\pi)$ spectra are multplied by $0.2$,
$\epsilon$$=$$0.2J$.
\label{fig4}}
\figure{SCF for $2$ mobile and static holes.
The $(\pi,\pi)$ spectra are
multiplied by $0.2$, $\epsilon$$=$$0.5J$.
\label{fig5}}
\begin{table}
\caption{(a) $\Delta_{\bbox{q}}(\bbox{k})$
for spin and charge operator with
$\bbox{q}$$=$$(\pi,\pi)$ ($16$ sites, $2$ holes, $t/J=2.5$).
(b) $n_{shift}(\bbox{q})$ for the same system.}
\begin{tabular}{l| c c  c c c c}
$\bbox{k},\bbox{q}$ & $(0,0)$ & $(\pi/2,0)$ & $(\pi,0)$ &
$(\pi/2,\pi/2)$ & $(\pi,\pi/2)$ & $(\pi,\pi)$ \\
\hline
(a), $O_s$ & -0.001 & -0.004 & -0.037 & +0.001 & +0.017 & +0.016 \\
(a), $O_d$ & -0.130 & -0.114 & +0.014& 0.000   & +0.111 & +0.114\\
\hline
(b), $O_s$ & 0.000  & 0.3174 & 0.2826 & 0.2863 & 0.1930 & 0.1778 \\
(b), $O_d$ & 0.000  & 0.7437 & 0.7911 & 1.0568 & 1.1673 & 1.1714 \\
\end{tabular}
\label{tab1}
\end{table}
\begin{table}
\caption{Hole occupation numbers $\bar{n}(\bbox{k})$
divided by $n_h$ for different $\bbox{k}$ and $n_h$
($16$ sites, $t/J=2.5$).}

\begin{tabular}{l| c c c c}
$n_h$ & $1$ & $2$ & $3$ & $4$ \\
\hline
$\bar{n}(\pi,\pi)/n_h$ & 0.1679 & 0.1567 & 0.1549 & 0.1460 \\
$\bar{n}(\pi,\pi/2)/n_h$ & 0.1406 & 0.1537 & 0.1427 & 0.1373
\end{tabular}
\label{tab2}
\end{table}

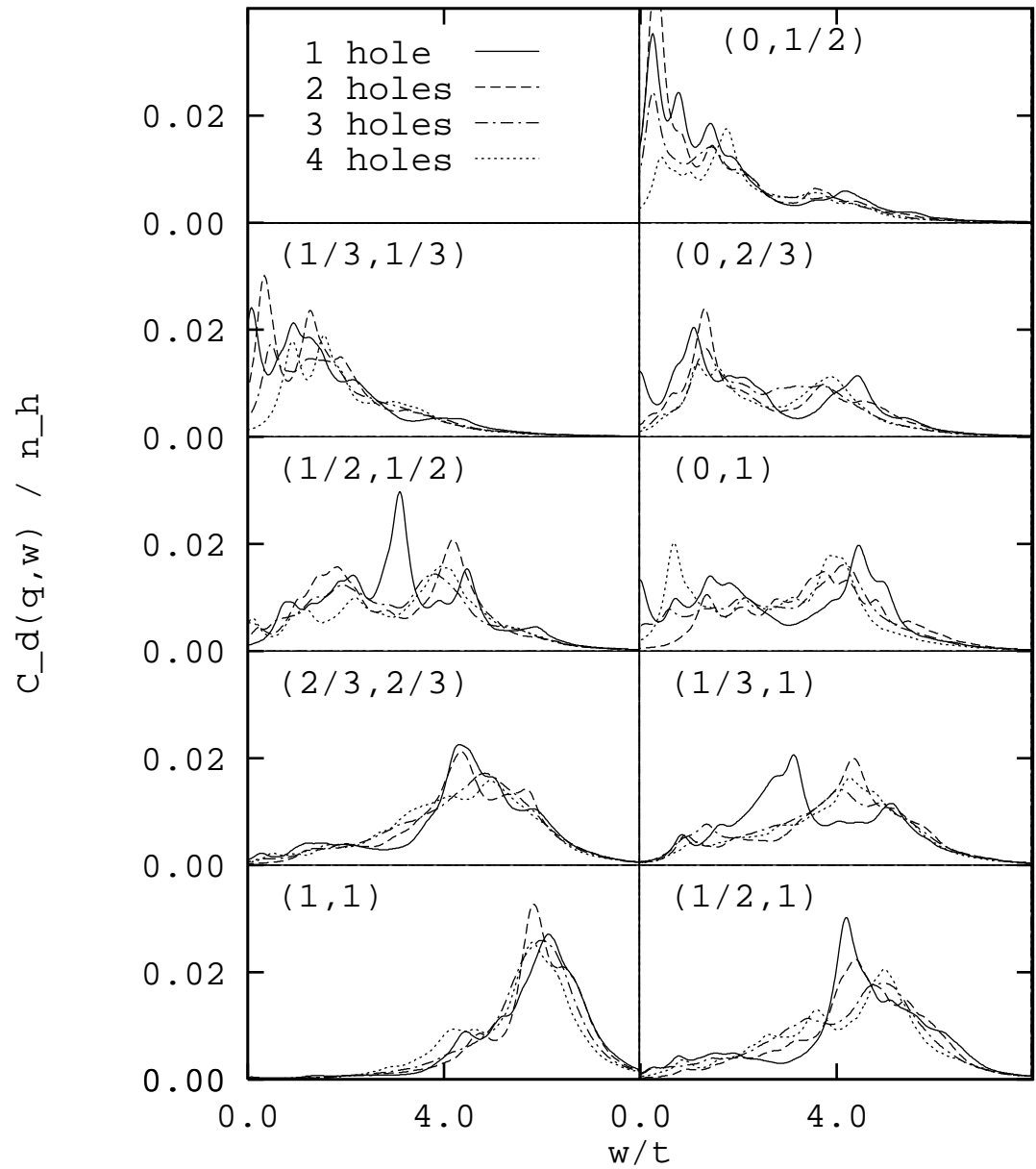
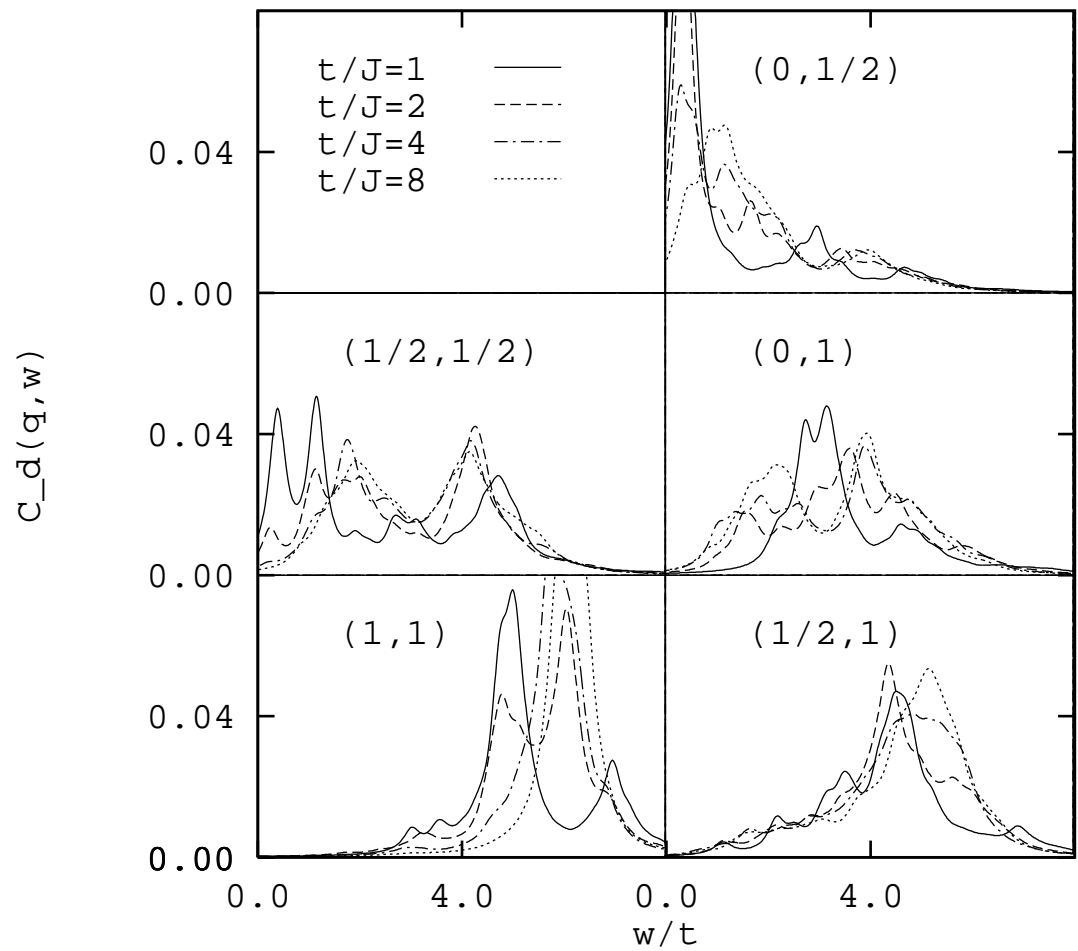
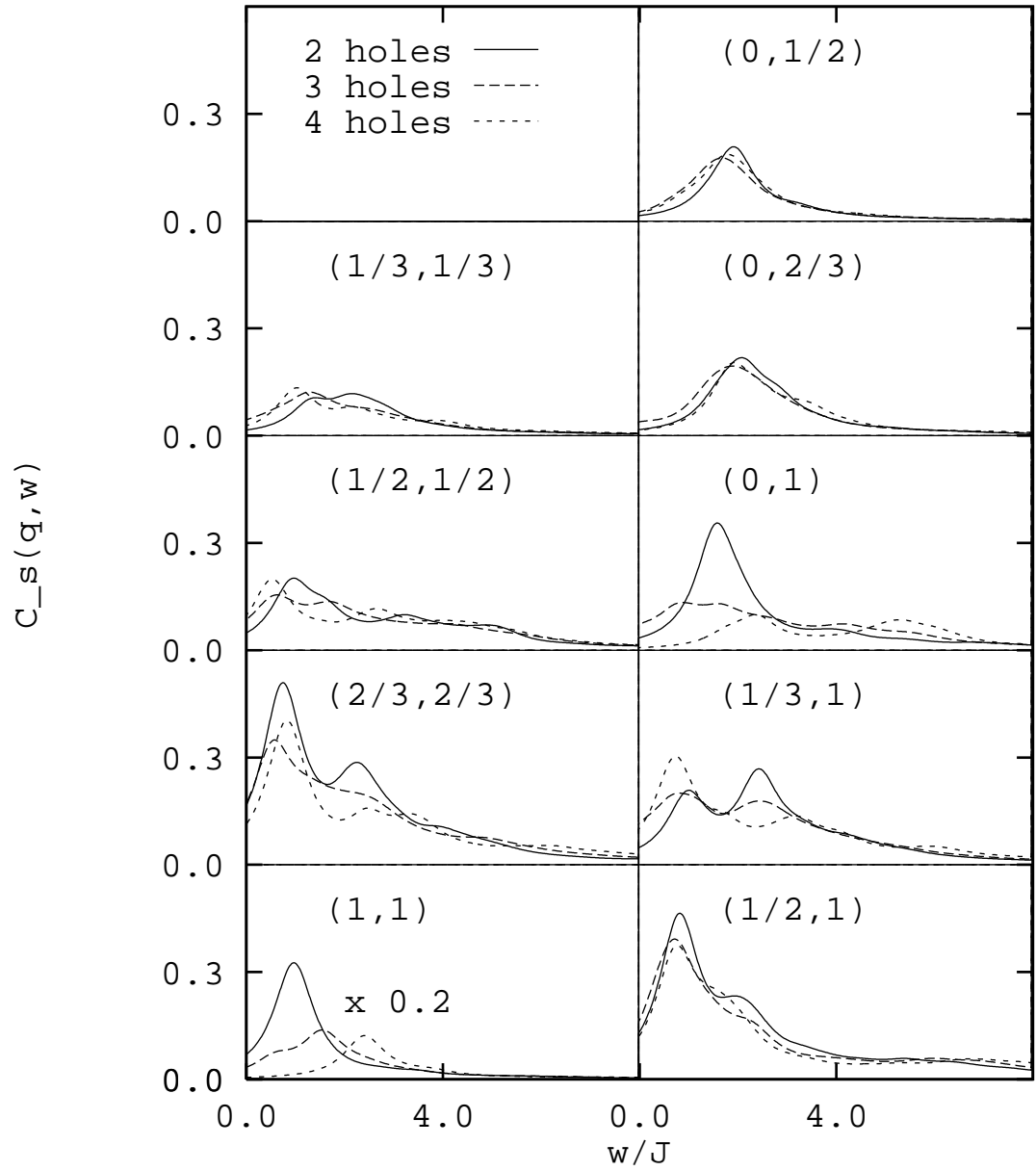
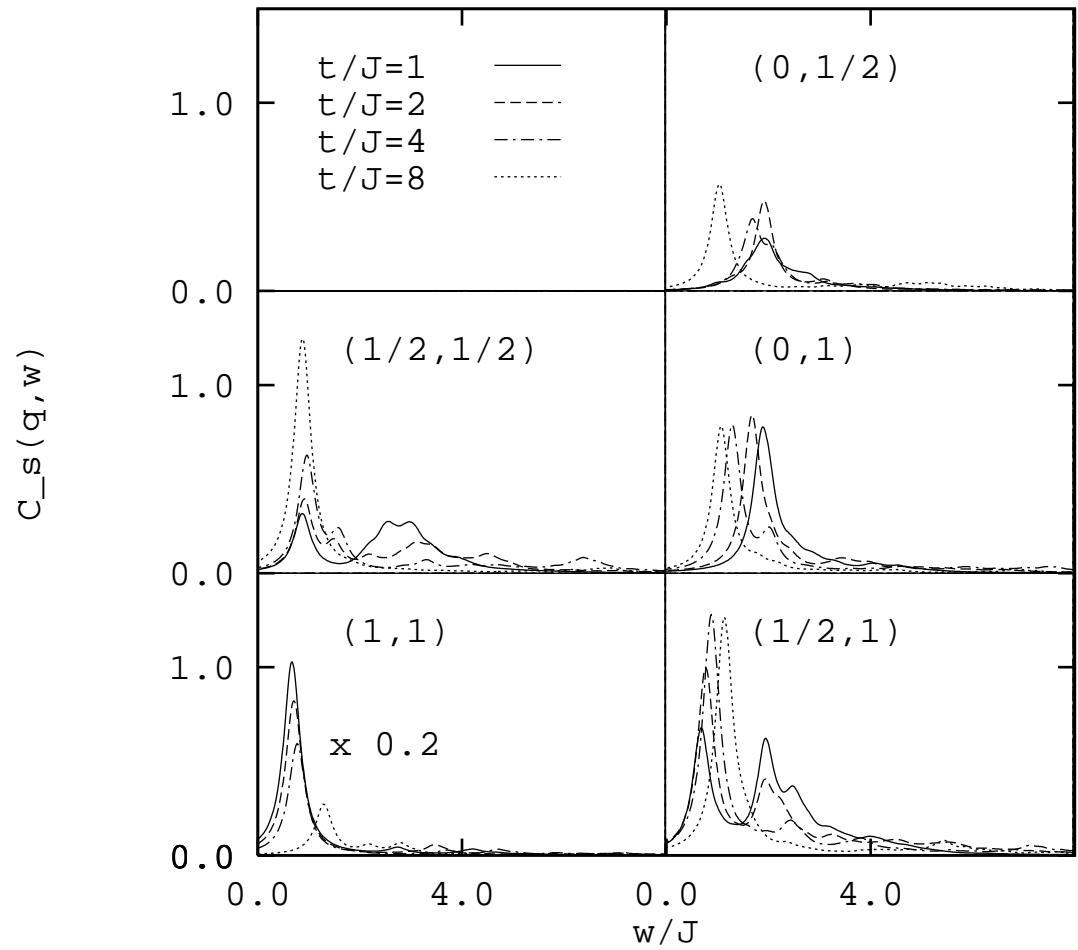
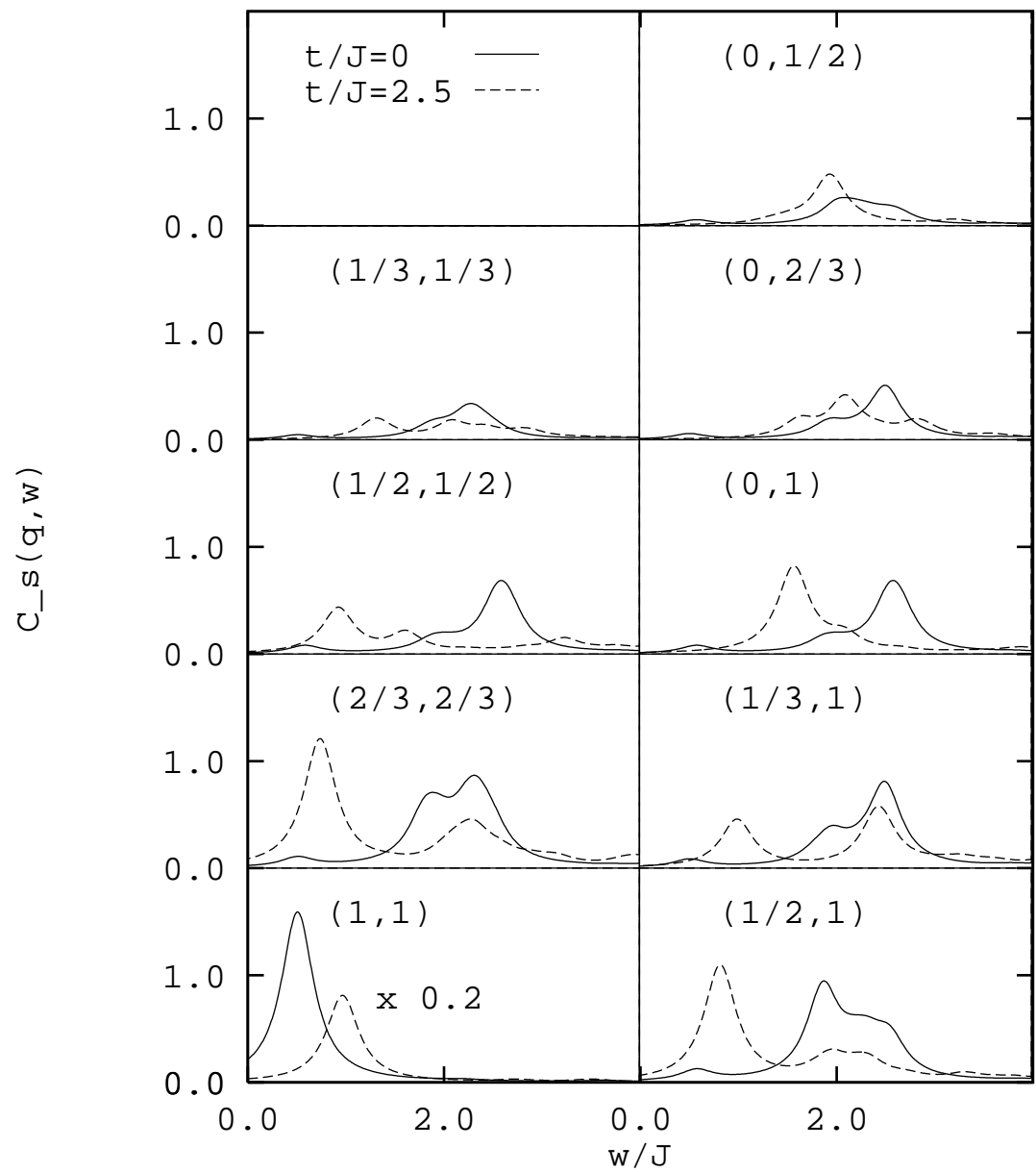
\end{document}